\documentclass[%
 reprint,
 amsmath,amssymb,
 aps,
]{revtex4-2}

\usepackage{graphicx}
\usepackage{dcolumn}
\usepackage{physics}
\usepackage{braket}
\usepackage{bm}

\begin{document}

\preprint{APS/123-QED}

\title{Stimulated emission tomography for efficient characterization of spatial entanglement}

\author{Yang Xu}
\affiliation{
Department of Physics and Astronomy, University of Rochester, Rochester, New York 14627, USA}
\email{yxu100@ur.rochester.edu}

\author{Saumya Choudhary}%
\affiliation{%
 The Institute of Optics, University of Rochester, Rochester, New York 14627, USA}%

\author{Robert W. Boyd}
 \affiliation{Department of Physics and Astronomy, University of Rochester, Rochester, New York 14627, USA}
\affiliation{
 The Institute of Optics, University of Rochester, Rochester, New York 14627, USA}%
 \affiliation{
 Department of Physics, University of Ottawa, Ottawa, Ontario K1N 6N5, Canada}%

\date{\today}

\begin{abstract}

Stimulated emission tomography (SET) is an excellent tool for characterizing the process of spontaneous parametric down-conversion (SPDC), which is commonly used to create pairs of entangled photons for use in quantum information protocols. The use of stimulated emission increases the average number of detected photons by several orders of magnitude compared to the spontaneous process. In a SET measurement, the parametric down-conversion is seeded by an intense signal field prepared with specified mode properties rather than by broadband multi-modal vacuum fluctuations, as is the case for the spontaneous process. The SET process generates an intense idler field in a mode that is the complex conjugate to the signal mode. In this work we use SET to estimate the joint spatial mode distribution (JSMD) in the Laguerre-Gaussian (LG) basis of the two photons of an entangled photon pair. The pair is produced by parametric down-conversion in a beta barium borate (BBO) crystal with type-II phase matching pumped at a wavelength of 405 nm along with a 780-nm seed signal beam prepared in a variety of LG modes to generate an 842-nm idler beam of which the spatial mode distribution is measured. We observe strong idler production and good agreement with the theoretical prediction of its spatial mode distribution. Our experimental procedure should enable the efficient determination of the photon-pair wavefunctions produced by low-brightness SPDC sources and the characterization of high-dimensional entangled-photon pairs.

\end{abstract}

\maketitle


\textit{Introduction.}\textbf{---}Entangled photon pairs have become an essential part of many quantum imaging \cite{PhysRevLett.92.033601, PhysRevLett.123.143603,kolobov2007quantum} and quantum information experiments \cite{nielsen2010quantum, murao2000quantum}. Recent studies have shown that high-dimensional entanglement has unique advantages in developing secure quantum communication protocols \cite{sec}, demonstrating the violation of Bell's inequality \cite{jack2010entanglement, aiello2005nonlocality}, and substantiating the EPR paradox \cite{leach2010quantum}. It was shown that the spatial mode of photons, such as the Lagurre-Gaussian (LG) modes, can be a convenient and natural choice of a high-dimensional basis \cite{erhard2020advances}. Consequently, the spatial entanglement of photon pairs has become a strong candidate for the implementation of high-dimensional quantum information protocols \cite{erhard2018twisted, ren2017spatially}. In recent years, there has been a plethora of research efforts, both theoretical \cite{spec,osorio2008correlations,torres2003quantum} and experimental \cite{interfere, beltran2017orbital}, for the accurate and efficient characterization of high-dimensional spatial mode entanglement in spontaneous parametric down-conversion (SPDC). 

In general, photon pairs produced by SPDC are entangled. The spatial structure of the down-converted photons produced in SPDC can be written in terms of a mode decomposition of their two-photon wave function in a complete set of spatial mode bases. Usually, the spatial modes of photons are decomposed into Laguerre-Gaussian (LG) modes, $LG_p^l$, where $l$ represents the orbital angular momentum (OAM), $l\hbar$, carried by the photon, and $p$ stands for the radial mode index. Measuring the probability amplitude for each basis mode is the core task in characterizing high-dimensional spatial mode entanglement. So far, there have been extensive research efforts \cite{krenn2017orbital, offer2018spiral, kulkarni2018angular, interfere} on the measurement of the two-photon entanglement in OAM basis, which gives rise to the spiral bandwidth. In these works, the use of image-rotating interferometers \cite{interfere, offer2018spiral} and the traditional coincidence counting method \cite{mair2001entanglement} are the two common ways to estimate the OAM mode distribution. Because of the low efficiency of SPDC processes, these two methods usually require painstaking alignment, a perturbation-free environment, and long data acquisition times. Furthermore, the SPDC experiments so far are capable of measuring the OAM ($l$) correlation of entangled photon pairs, wherein the contributions from individual radial modes ($p$) are traced out  \cite{kulkarni2018angular}. 

However, the radial mode ($p$) is also an important quantum number describing the radial-momentum-like property of LG modes. The entanglement in radial modes offers a larger number of qubits that can be used in quantum communication \cite{pang2018demonstration}. In addition, the use of radial modes is essential in controlling the structures of topological knots of phase and polarization singularities \cite{larocque2018reconstructing}. These knot structures can be generated by a specific coherent superposition of LG modes \cite{leach2005vortex} and can be robust against turbulence \cite{dennis2010isolated}. This feature may become useful in the propagation of entangled photons through turbulent media. A full modal analysis, including the radial modes, of the two-photon spatial mode entanglement produced by SPDC has been worked out theoretically \cite{spec}. In principle, these theoretical results can be verified by flattening the phase of the down-converted photons and subsequently performing coincidence measurements through single-mode fiber-coupled single photon detectors (SPDs). However, the low photon pair generation rate along with the inherent photon losses associated with projective measurements makes the experimental realization particularly challenging. 

Fortunately, a recently developed technique of stimulated emission tomography (SET) \cite{set}, which involves the use of stimulated parametric down-conversion to characterize the photon pairs produced during SPDC, is a promising alternative. By exploiting the link between spontaneous and stimulated processes, SET offers an efficient way to estimate the quantum state of the photon pair with a signal-to-noise ratio several orders of magnitude larger than the traditional quantum state tomography method \cite{yang2020stimulated, fang2016multidimensional}. This large enhancement in signal-to-noise ratio allows the use of traditional detectors in the measurement of the joint spectral density of entangled photon pairs \cite{rozema2015characterizing, keller2022reconstructing, xu2023orthogonal}. In this Letter, we demonstrate a new way to efficiently measure the entanglement of photon pairs in the Laguerre-Gaussian (LG) basis employing the SET technique. We seed the down-conversion process with an $LG^l_0$ beam and measure the intensity of the dominant LG modes in the generated idler to obtain the idler's spatial mode distribution for a particular seed. We repeat this process for seed beams of several dominant $LG^l_0$ modes that can be produced spontaneously during the down-conversion process to reconstruct the joint spatial mode distribution (JSMD) of the photon pairs in the LG basis.

\textit{Theory.}\textbf{---}Consider a typical SPDC setup that consists of a c.w. Gaussian pump beam with frequency $\omega_p$ propagating along the $z$ direction incident on a thin nonlinear crystal of length $L$. The output quantum state $\ket{\psi}$ of the photon pair leaving the nonlinear crystal in the momentum domain is given by \cite{walborn2010spatial} 
\begin{align}
    \ket{\psi} = \iint d\mathbf{k}_sd\mathbf{k}_i \Phi(\mathbf{k}_s,\mathbf{k}_i)\hat{a}^\dag_s(\mathbf{k}_s)\hat{a}^\dag_i(\mathbf{k}_i)\ket{0}
\end{align}
where $\Phi(\mathbf{k}_s,\mathbf{k}_i)$ is the photon pair wave function describing the pump and the phase matching conditions, $\hat{a}^\dag_s(\mathbf{k}_s)$, $\hat{a}^\dag_i(\mathbf{k}_i)$ are creation operators for the signal and idler modes with wave vectors $\mathbf{k}_s$ and $\mathbf{k}_i$. The subscripts $p, s, i$ refer to the pump, signal, and idler waves respectively. To characterize the spatial mode entanglement of an SPDC source, we need to calculate the JSMD, $\left|C_{p_s,p_i}^{l_s,l_i}\right|^2\equiv \left| \braket{l_s,p_s,l_i,p_i|\psi}\right|^2$ where $\ket{l_s,p_s,l_i,p_i}$ is the output state in which the signal and idler photon are found in $LG_{p_s}^{l_s}$ and $LG_{p_i}^{l_i}$ mode. The probability amplitude $C_{p_s,p_i}^{l_s,l_i} $ is determined by the integral
\begin{align}
    C_{p_s,p_i}^{l_s,l_i} = \iint d\mathbf{k}_sd\mathbf{k}_i \Phi(\mathbf{k}_s,\mathbf{k}_i)\left[LG_{p_s}^{l_s}(\mathbf{k}_s)\right]^* \left[LG_{p_i}^{l_i}(\mathbf{k}_i)\right]^*.
    \label{eqn:oam}
\end{align}
The value $\left|C_{p_s,p_i}^{l_s,l_i}\right|^2$ represents the probability of finding the down-converted photon pair in the state $\ket{l_s,p_s,l_i,p_i}$. 
Assuming a monochromatic Gaussian pump and a thin crystal, $C_{p_s,p_i}^{l_s,l_i}$ becomes \cite{supp, spec}: 

\begin{align}\label{eq:oam_amp}
   C_{p_s,p_i}^{l,-l}& \propto A^{\abs{l}}_{p_s,p_i} \frac{\left(1-\gamma_s^2 + \gamma_i^2\right)^{p_s} \left(1+\gamma_s^2 - \gamma_i^2\right)^{p_i} (-2\gamma_s\gamma_i)^{\abs{l}}}{(1+\gamma_s^2 + \gamma_i^2)^{p_s + p_s + \abs{l}}} \notag\\
   &\times {}_2F_1\left(-p_i,-p_s;-p_i-p_s-\abs{l}; \frac{1-(\gamma_s^2+\gamma_i^2)^2}{1-(\gamma_s^2-\gamma_i^2)^2}\right)
\end{align}
where $A^{\abs{l}}_{p_s,p_i} = {(p_s+p_i + \abs{l})!}/{\sqrt{p_s!p_i!(p_s+\abs{l})!(p_i+\abs{l})!}}$, ${}_2F_1(\boldsymbol{\cdot},\boldsymbol{\cdot};\boldsymbol{\cdot};\boldsymbol{\cdot})$ is the Gaussian hypergeometric function, $\gamma_i = w_p/w_i$, and $\gamma_s = w_p/w_s$ are the inverses of the signal and idler beam waist diameters normalized to the waist diameter of the pump beam, respectively. The conservation of angular momentum $l \equiv l_s = -l_i$ is guaranteed by the integrals in Eq. (\ref{eqn:oam}) for thin crystals.

In most common setups of SPDC, we have $w_i = w_s$ leading to $\gamma_i = \gamma_s = \gamma$. For experimental and computational simplicity, we only study the case where $p_s = p_i = 0$ in our experiment. Thus, the expected coincidence probability simplifies to \cite{supp}
\begin{align}\label{eq:spec}
    \abs{C_{0,0}^{l,-l}}^2 \propto \left(\frac{2\gamma^2}{1+2\gamma^2}\right)^{2\abs{l}}
\end{align}


\begin{figure}[hbt!]
  \centering
  \includegraphics[width=0.5\textwidth]{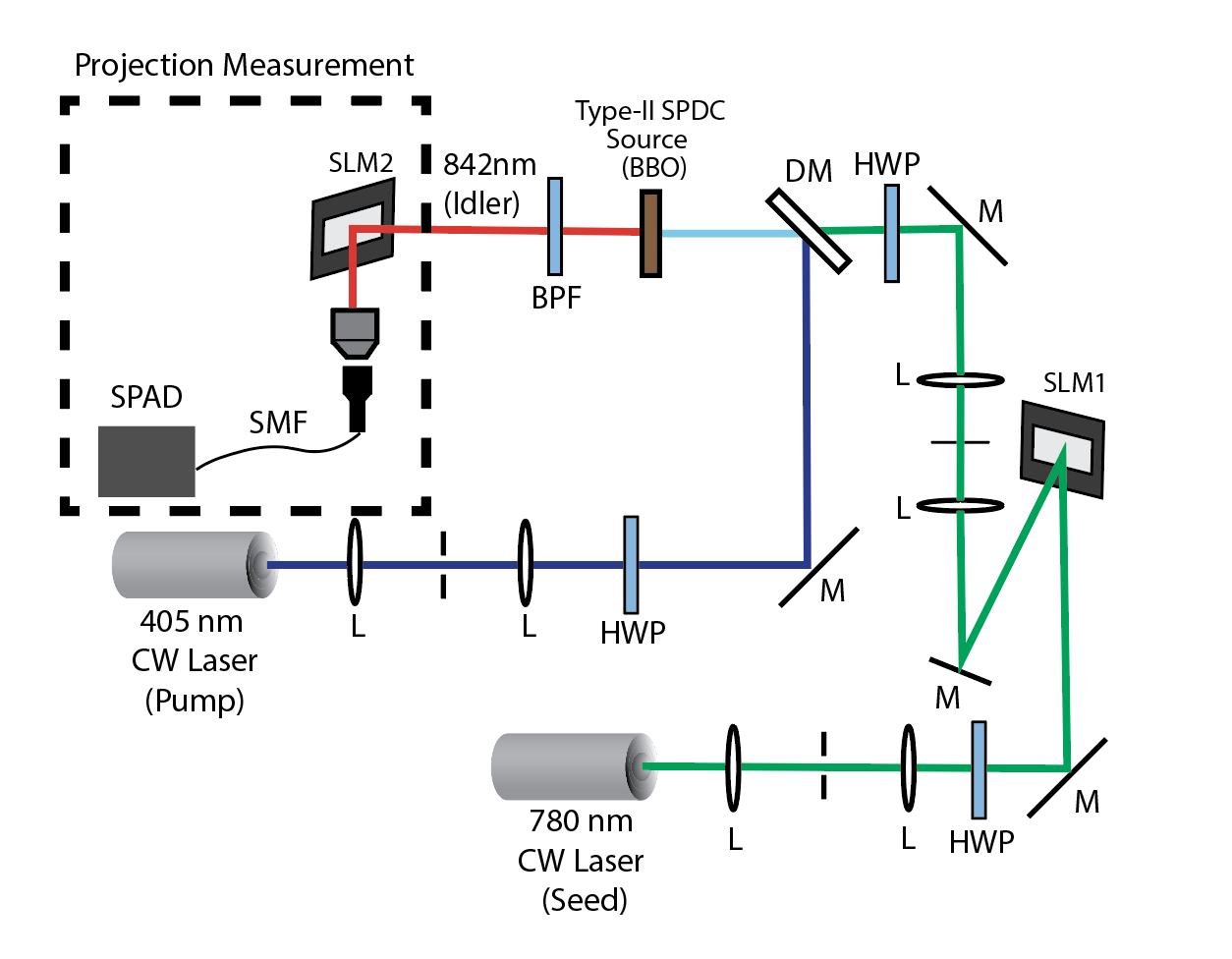}
\caption{Schematic of the experimental setup. A type-II BBO crystal is pumped by a 405-nm c.w. laser beam with a Gaussian spatial profile (drawn in blue). The seed is generated in the first diffraction order of a holograph encoded on a spatial light modulator (SLM1). The seed is injected into the type-II BBO crystal together with the Gaussian pump beam. The output idler beam is directed to the spatial mode projection measurement setup where the intensity of the phase-flattened beam is measured to construct the JSMD. L: lens; HWP: half-wave plate; BPF: band-pass filter; SMF: single-mode fiber; DM: dichroic mirror.}
\label{fig:setup}
\end{figure}

\textit{Experiment.}\textbf{---}Figure \ref{fig:setup} shows the SET setup used to measure the LG mode spectrum with SET. We pump a 2-mm-long type-II beta barium borate (BBO) crystal with a horizontally polarized and collimated Gaussian ($LG_0^0$) beam at a wavelength of 405 nm. The pump beam has a waist diameter of 2 mm and a power of 20 mW. To generate the seed beam in the $LG^{l_s}_0$ mode, we isolate the first diffraction order of a collimated c.w. Gaussian beam at 780 nm modulated by the appropriate diffractive hologram realized on a spatial light modulator (Meadowlark E-series 1920 $\times$ 1200 SLM) (SLM 1) \cite{Arrizon}.
The prepared seed beam in the Laguerre-Gaussian mode $\text{LG}_0^{l_s}$ is then injected into the BBO nonlinear crystal together with the Gaussian pump beam at 405 nm. We spectrally filter both the pump and seed beams with narrowband filters (10 nm full-width at half-maximum) centered at 405 nm and 780 nm, respectively, and subsequently adjust their polarizations using separate half-wave plates to satisfy the phase-matching condition for the crystal. By aligning the seed beam with the SPDC signal beam of the same polarization at 842 nm, the idler beam of the opposite OAM number $l_i = -l_s$ is generated. A narrowband filter centered at a wavelength of 840 nm isolates the idler. We then use the SLM2 (Santec SLM-200) to flatten the phase of the filtered idler beam by applying the appropriate phase holograms  \cite{miyamoto2011detection,nicolas2015quantum}. The phase-flattened beam couples to a single-mode fiber (SMF). We use a single photon avalanche diode (SPAD) to measure the total number of photons coupled into the SMF. 


\textit{Results.}\textbf{---}We first verify the validity of the thin-crystal approximation in our experiment. The 405-nm Gaussian pump beam we use in our experiment has a waist diameter of 2 mm. The thin-crystal approximation for this pump beam is valid 
when $L \ll z_R$, where $L$ is the length of the crystal and $z_R$ is the Rayleigh range of the pump beam. This condition is equivalent to $w_p/\sqrt{\lambda_p L} \gg 1$ \cite{PhysRevA.103.063508}. In our experiment, the length of our type-II BBO crystal is $L=2.0$ mm, so the quantity $w_p/\sqrt{\lambda_p L} \approx 94.8 $ is much greater than $ 1 $. Thus, the thin-crystal approximation is valid for our experiment.  

We measure the number of photons collected by the SPAD over ten 5-second time windows for different seed modes $(LG^{l_s}_0)$ ranging from $l_s=-6$ to $l_s=6$. For all measurements, we subtract the average dark counts of the SPAD, which was determined by blocking all the beams and measuring the photon number integrated within a five-second window averaged over twenty trials.

\begin{figure}[hbt!]
  \centering
  \includegraphics[width=0.5\textwidth]{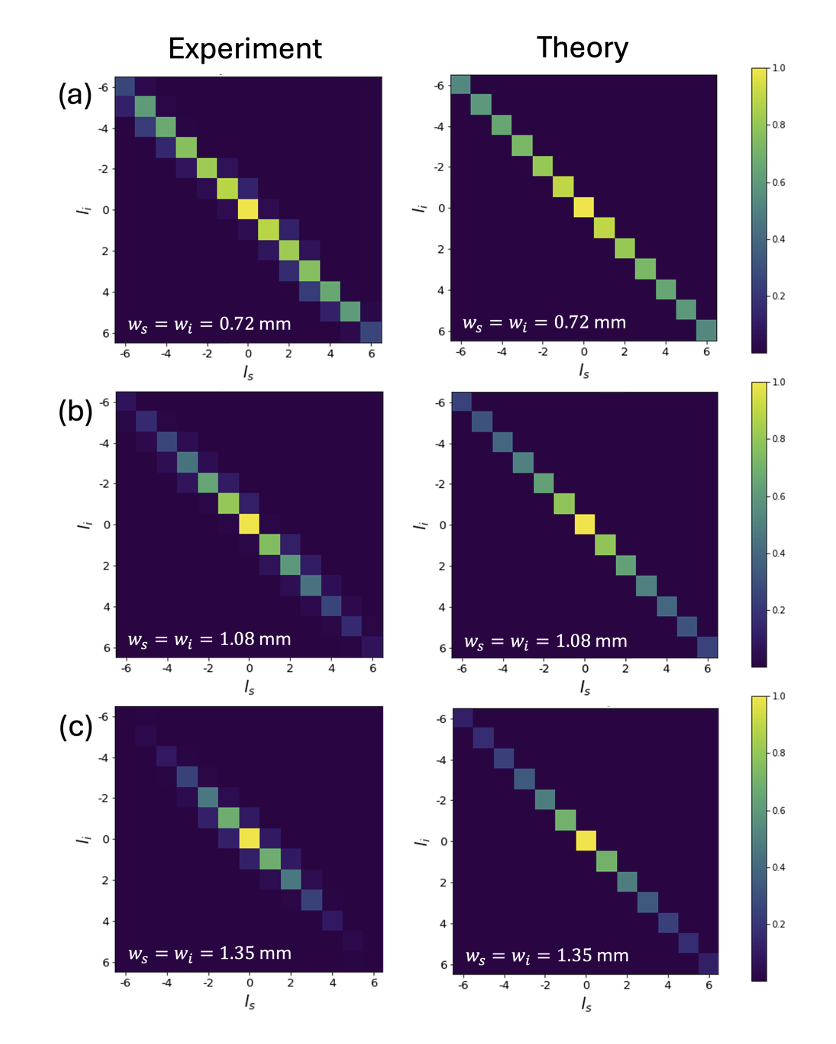}
    \caption{Two-photon JSMD measured by SET compared to theoretical predictions. Projective measurements are performed on the idler beam for different LG modes while keeping the signal spatial mode fixed. The process is repeated for all of the LG signal modes. The measured idler photon counts are normalized to the overall maximum of the idler modes to produce the JSMD. The width of the pump is kept constant at 2 mm. Experimental measurements agree well with the theoretical results. Note that the width of the distribution decreases as the signal and idler waist diameters are increased. (a) JSMD for $w_i = w_s = 0.72$ mm (b) JSMD for $w_i = w_s = 1.08$ mm (c) JSMD for $w_i = w_s = 1.35$ mm.}
\label{fig:result_mat}
\end{figure}

In Fig. \ref{fig:result_mat}, we show the experimentally measured (left panels) and the theoretically calculated (right panels) JSMD matrices for a pump width of 2 mm and signal and idler widths of (a) 0.72 mm, (b) 1.08 mm and (c) 1.35 mm. The theoretical results show the JSMD $|C^{l, -l}_{0, 0}|^2$ defined in Eq. (\ref{eq:oam_amp}) calculated for the corresponding pump, signal, and idler. The two-dimensional (2D) JSMD is measured experimentally in the aforementioned way. We notice that only the blocks where $l_i+l_s=0$ is satisfied are significant given a Gaussian pump beam $(LG^0_0)$ at 405 nm. The measured two-photon JSMD agrees with theoretical calculations and thus confirms the conservation of OAM in collinear SPDC processes. 

\begin{figure*}[hbt!]
  \centering
  \includegraphics[width=0.95\textwidth]{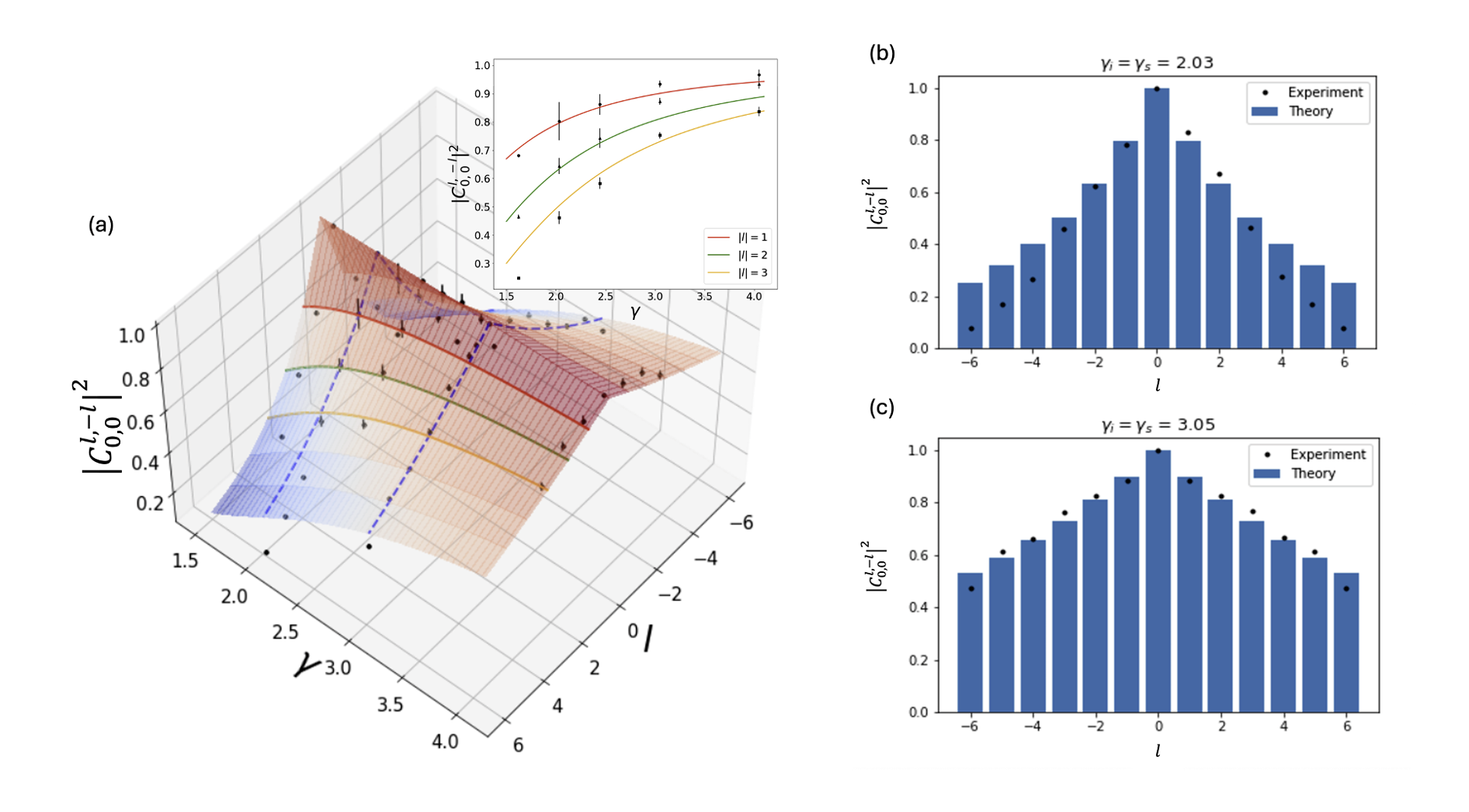}
\caption{(a) Spatial mode distribution as a function of the azimuthal index $(l)$ and the normalized seed beam waist $(\gamma)$. The colored surface shows the theoretical values. The LG mode spectra for two different signal beam waist diameters (dashed blue line cuts) are shown in detail in (b) and (c). The inset of (a) shows the dependence of the LG mode distribution on the normalized seed beam waist for $l_s = -l_i = 1, 2, 3$ where $\gamma = \gamma_s = \gamma_i = w_p / w_{s,i}$. The inset plot shows that the weight of the LG mode decreases for higher LG modes and for larger signal-to-idler beam waist ratios.  (b) LG mode distribution at $\gamma_i = \gamma_s = 2.03$; (c) LG mode distribution at $\gamma_i = \gamma_s =3.05$. Note $\gamma_m = w_p/w_m$ where $m = s,i$. (b) and (c) shows the LG mode distribution of the idler measured by SET compared to theoretical predictions. The beam width of the output idler is the same as the seed width due to the nature of a stimulated process. The distribution widens for smaller widths of the seed beam due to a larger spatial overlap of the pump beam with seed beams in LG modes with higher OAMs.}
\label{fig:result_spec}
\end{figure*}

From Eq. (\ref{eq:spec}), the spatial mode distribution depends on both the azimuthal index $(l)$ and the normalized beam waist $(\gamma)$. Figure \ref{fig:result_spec}(a) shows the theoretical and experimentally measured spatial mode distribution for the dominant LG modes and multiple beam waist diameters. The colored surface represents the theoretical predictions given by Eq. (\ref{eq:spec}). We first examine the LG mode spectrum of the down-converted photon when the beam waist diameter is specified. Figure \ref{fig:result_spec}(b) and Fig. \ref{fig:result_spec}(c) shows the LG mode probability distribution $|C^{l, -l}_{0, 0}|^2$ for two different signal seed beam waist diameters in comparison with theoretical calculations. In Fig. \ref{fig:result_spec}(b) where the seed width is larger($\gamma_{s} = \gamma_i = 2.03$), there is a noticeable discrepancy between theoretical predictions and the measured data for higher LG modes. This effect is a consequence of reduced coupling efficiency of the projected idler mode to the SMF for higher values of $l$ at larger idler beam waist diameters. SLM2 flattens the phase into a wider beam. Thus, for each spatial mode projection phase mask on SLM2, the coupling efficiency to the fiber varies slightly. This deviation from the optimal fiber coupling becomes obvious for smaller $\gamma_s$ (larger $w_s$) since the width of the converted beam starts to exceed the field of view of the objective we use in our fiber coupling system. Fortunately, this problem could be remedied using an objective with a larger field of view. 

In Fig. \ref{fig:result_spec}(b) and Fig. \ref{fig:result_spec}(c), we notice that the LG mode distribution widens as the waist of the seed beam $w_s$ (and $w_i$) is reduced, (or equivalently $\gamma_s$ and $\gamma_i$ is increased). From Eq. \ref{eqn:oam}, the weight of each LG mode should depend on the ratio between the pump and the signal/idler widths, $\gamma_{s,i} = w_p/w_{s,i}$. In stimulated parametric down-conversion, the relation $\gamma_s = \gamma_i$ is automatically satisfied. Hence, the probability amplitude of each LG mode only depends on the normalized beam waist $\gamma = \gamma_s = \gamma_i$. To quantify the dependence of the width of the LG mode distribution on the waist parameter $\gamma$, we show the normalized coincidence probability of the photon pair production, $\left|C_{0,0}^{l,-l}\right|^2$, as a function of $\gamma$, for three different values of $l_i$ and $l_s$ with $l_i + l_s = 0$ in the inset of Fig. \ref{fig:result_spec}(a). We note that the coincidence probability increases for all values of $l$ as the waists of the signal and idler decrease ($\gamma = \gamma_i = \gamma_s$ increases). This broadening in the width of the distribution can be explained as follows: a smaller seed beam waist allows higher LG modes to have spatial overlap with a fixed pump beam. This results in a higher numerical value of the overlap integral between the two electric field's spatial profiles.

\textit{Discussion.}\textbf{---} In order to show the improvement in efficiency improvement with our SET method, we examine the SPDC process where the Gaussian pump is down-converted into a Gaussian signal and idler beams $(p_p,l_p) = (p_s,l_s) = (p_i,l_i) = (0,0)$ as a benchmark case. We compare the photon pair detection efficiency measured in one of our earlier experiments \cite{PhysRevLett.123.143603} performed on the same pump laser with the idler photon detection efficiency measured by SET.

In our earlier coincidence counting setup, the signal and idler beams were coupled to single-mode fibers, and the coincidence counts were measured by single photon detectors and a coincidence-counting circuit with a time window of  $\sim13$ ns. We measured a detection efficiency of $\sim$ $5.0 \times 10^{6}$ $\text{Hz}/\text{W}/\text{m}$ for type-II BBO. In comparison, in our SET experiment with the same pump laser, we measure a photon detection efficiency of $\sim 1.6 \times 10^9$ $\text{Hz}/\text{W}/\text{m}$ for type-II BBO with a seed beam of $\sim$ 9 mW. This orders-of-magnitude improvement suggests that our technique based on SET offers a time-saving method to fully characterize low-brightness broadband sources of entangled photons such as novel nonlinear metasurfaces \cite{santiago2021photon}. These sources have much lower photon pair generation rates because of their tiny interaction lengths. Another potential advantage of our experimental scheme is the aforementioned ability to efficiently measure the full modal structure of the entangled photon pairs produced by SPDC, including both radial and angular parts of LG modes. Our method can be applied to directly measure the full two-photon spatial mode spectrum, including both the angular and the radial modes, with a modest modification. To achieve this, we can simply replace the projection measurement unit in our experiment setup with the recently developed radial mode sorters \cite{zhou2017sorting} or the latest multiplane light conversion method \cite{choudhary2018measurement, zhang2023multi} that can sort the spatial mode of the incident beam into an arbitrary set of spatial modes. 

\textit{Conclusions.}\textbf{---}We report the first demonstration of an efficient classical method to estimate the full JSMD of entangled photon pairs in the high-dimensional space produced by an SPDC source. We estimate the JSMD for the lowest radial modes ($p_i = p_s =0$) and study its dependence on beam waist diameters in a complete set of Laguerre-Gaussian spatial modes ($\text{LG}_l^p$). We demonstrate that our method can be easily adapted into any spatial mode basis with the current spatial modulation techniques. Our experimental demonstration of the SET technique for spatial modes can be used for a full characterization of any weak down-conversion sources such as low-brightness sources based on nonlinear metasurfaces \cite{santiago2021photon}. 

\bibliography{main}

\end{document}